\documentclass[cits]{PoS}
\usepackage{amsmath}
\usepackage{multirow}
\title{Lattice Kaon Physics}

\ShortTitle{Lattice Kaon Physics}

\author{\speaker{P A Boyle}\thanks{University of Edinburgh}\\
        School of Physics\\
	SUPA\\
	JCMB\\
	University of Edinburgh\\
	Edinburgh UK EH9 3JZ\\
        E-mail: \email{paboyle@ph.ed.ac.uk}}

\def\mev{\,\mathrm{MeV}}
\def\fm{\,\mathrm{fm}}
\def\SU{\mathrm{SU}}
\def\su#1#2{\SU(#1)_\mathrm{#2}}

\def\rpisqsu2sim{0.354(31)}  

\newcommand{\tchpt}{$\chi^{\rm PT}$}

\abstract{I review lattice quantum chromodynamics as relevant to the kaon system.
Topics covered include the pseudoscalar masses and decay constants, 
the chiral effective lagrangian, $f_K/f_\pi$,  semi-leptonic kaon decay form
factors, and the neutral kaon oscillation parameter $B_K$.
}

\FullConference{2009 KAON International Conference KAON09,\\
		 June 09 - 12 2009\\
		 Tsukuba, Japan}

\begin{document}

This paper reviews the status of Lattice QCD where relevant to kaon physics.
Some summary intended for non-experts of the broader status of the field is 
given in section~\ref{Sec:StateOfTheNation}.
I interpret this goal as including some aspects of pion
physics and the Chiral Lagrangian in section~\ref{Sec:ChiralLagrangian}.

I review the status of Lattice QCD input to determining $V_{us}$, section~\ref{Sec:Vus}, 
both via $f_K/f_\pi$, section~\ref{Sec:FKFPI},  
and $K_{l3}$ section~\ref{Sec:KL3}. 
The kaon bag parameter $B_K$ which contributes to the $\epsilon_K$
constraint on the unitarity triangle is reviewed, section~\ref{Sec:BK}. 
Lattice $K-\pi\pi$ calculations were reviewed at this conference by Norman 
Christ \cite{Christ} and are thus beyond the scope of this review.

Carrying out a review of this nature is an increasingly daunting task
in the presence of the excellent Flavia Lattice Averaging Group effort
\cite{FLAG}.
It is also worth mentioning excellent recent reviews
by individual members of FLAG 
\cite{Lellouch:2009fg,Necco:2009cq}.
In many areas it is difficult to match
this comprehensive effort. My approach therefore is to try to highlight
theoretical issues, distinguishing features, and methods
of the lattice calculations that are not easily conveyed by
the compressed asterisk rating system; I do not necessarily attempt the 
comprehensive categorisation and averaging task this working group 
has undertaken
except where I may have significant updates from new material, or
wish to differ in the analysis.

PAB wishes to thank the many authors whose original research has been reviewed and
emphasizes that any use of Lattice world averages or quoted results herein must
cite the orginal source(s) where the intellectual value was created.

\section{Current Large Scale Lattice Simulations}
\label{Sec:StateOfTheNation}

Despite many recent advances in both algorithms and computing hardware
Lattice QCD cannot presently be performed at physical masses in a 
large volume for reasons of numerical cost. Were this feasible, there
would be little debate about methods in the field. 
In practice, the choice of approach, table~\ref{tab:Sims}, is a complex
optimisation problem where the selected approach is dependent on
the physical processes for which one aims to yield a minimal 
overall error. These errors are composed of statistical and
finite volume, mass extrapolation and discretisation systematics.
Strategy is influenced by estimates 
of the importance \& calculability of finite volume, NNLO chiral
expansion effects, and even the influence of strange quark loops.

The physics goal of computing chirally structured weak matrix elements  
($B_K$ and $K\to \pi\pi$, for example) makes perhaps the
biggest rationally justifiable difference of approach. RBC-UKQCD and
JLQCD use substantially more expensive methods to obtain near
exact chiral symmetry and gain access to richer
phenomenology than would otherwise be possible. These approaches
also give automatic $O(a)$ improvement off-shell, making improved RI-mom non-perturbative
renormalisation simpler. However, less theoretically pristine approaches
are substantially cheaper and BMW and MILC in particular have used this to gain access 
to finer lattice spacings and lighter simulated masses in large volumes. 


\begin{table}[hbt]
{\footnotesize{
\begin{tabular}{c|c|c|c|c|c|c|c}
Collaboration  & Action & {L(fm)}&{$m_\pi^{\rm min}$ (MeV)}& {$a^{-1}$ GeV}& {ChPT}&{Unitary}& {FV corr. }\\
\hline

JLQCD\cite{Kaneko:2006pa}
    & Overlap& {1.7}& {310}& {1.8} & {$\su{2}{}$}& {U}& {CDH}\\
RBC-UKQCD \cite{Allton:2008pn}& DWF    & {2.0,2.7}&{330}&{1.7,2.3}& {$\su{2}{}$, $\su{3}{}$}&{PQ}&{-}\\
ETMC  
\cite{Blossier:2007vv}
   & Twisted mass &{2.0}& {300}&{1.9,2.2,2.8}&{$\su{2}{}$}& {U}&{CDH}\\
PACS-CS\cite{Aoki:2008sm}
 & Clover & {3.1}&{156}&{2.2}& {$\su{2}{}$, $\su{3}{}$}&{U}&{CDH}\\
BMW  \cite{Durr:2008zz}    & Clover &{4.0}&{190}& {1.6,2.3,3.03}&{polynomial}&{-}& {-}\\
\multirow{2}{*}{MILC\cite{Bazavov:2009bb}} & \multirow{2}{*}{Staggered} &  2.4,2.9 &\multirow{2}{*}{320}& 1.1,1.3,1.6 
                           &\multirow{2}{*}{rs-$\su{3}{}$}&\multirow{2}{*}{PQ}& \multirow{2}{*}{-}\\
                           &                            &    3.4                        &                    & 2.2,3.3,4.4& & 
\end{tabular}
}}

\caption{Summary of the parameters of major lattice calculations reviewed
\label{tab:Sims}
}
\end{table}

Lattice QCD gives complete
freedom in choice of the quark masses (dynamical loops and valence legs)
in numerical simulation. Purely unitary simulations 
maintain the valence quark masses equal to those of simulated
dynamical flavors in all matrix elements.
A \emph{partially quenched} simulation is a superset of the corresponding
unitary simulation; here the multiple valence masses are used
(both equal to and not equal to the dynamical masses)
and next-to-leading order PQ-\tchpt is available for a variety of important quantities
such as pseudoscalar masses, decay constants and the kaon bag parameter.

RBC-UKQCD and MILC have used partially quenched analyses to increase the information
within the chiral regime, while JLQCD, ETMC, BMW and PACS-CS used unitary simulations.

Where smaller volumes are used, it is common for finite volume corrections 
to be applied to the data prior to chiral extrapolation. 
This, of course, takes the effect of finite volumes as an input to 
(rather than output of) the lattice calculation.
The presently favoured model is the resummed scheme of Colangelo, Durr and Haefeli (CDH) and is 
applicable only to unitary datapoints, and cannot be combined with partially quenched analyses.
For this reason the calculations where $m_\pi L$ becomes rather smaller than around 3.2 
both apply finite volume correction and consider only unitary datapoints.

Approaches to mass extrapolations include the application of SU(3) and SU(2) 
chiral effective theories. BMW recently adopted Taylor expansion 
having achieved both light simulation masses and a large volume. They
feel sufficiently close to the physical point to include polynomial
fits in their analysis; these are viewed as an analytic expansion around the non-zero physical quark mass in
a massive region whose validity does not extend to the chiral non-analyticity at zero.

\subsection{Is lattice QCD \emph{ab initio} in practice?}

Lattice QCD has rightly enjoyed the reputation of an \emph{ab initio}
method to solve QCD. 
However, present analysis \& extrapolation approaches do
involve varying degrees of \emph{non}-lattice input. 
What is clear is that, in some cases, the extrapolated results 
have relinquished the significant benefit of having only the QCD Lagrangian as input.

Ideally, any use of the low energy expansion of QCD
for chiral extrapolation should be performed in a consistent
\& well motivated context. Either \emph{complete} NLO or 
\emph{complete} NNLO should be used; it should both describe the data
and appear a well convergent series within the range of the data
without ad hoc resummation schemes or partial NNLO where only favoured
analytic terms are included, or certain LEC's biased towards phenomenological input.

Taking the long view, lattice QCD remains systematically improvable.
Continued advance in supercomputing power will soon remove the
need for such extrapolations, just as the quenched approximation has been
long discarded. However, it is worth highlighting 
mass extrapolation and finite volume issues as areas where the field should do better.

\section{Chiral Effective Lagrangian}

\label{Sec:ChiralLagrangian}

The chiral effective theory can be viewed in two roles by a lattice
practitioner. First it can be viewed as an important predictive tool which
can be verified and understood with lattice QCD (i.e. intrinsic interest).
The chiral effective theory has also been used as 
a critical component of making well founded mass extrapolations from 
simulation in almost all recent lattice calculations
(i.e. practical value). The former requires that chiral perturbation theory
be a well convergent series at the relevant physical masses, while the latter 
requires convergence above the physical masses.

Were SU(3) chiral effective theory convergent at the kaon mass, it would
be the natural framework for interpreting lattice data. 
MILC in particular have relied on the partially quenched SU(3) approach 
to remove unitarity violating
discretisation effects introduced by their ``rooting'' prescription 
\cite{Aubin:2003rg}.
However, some determinations of the leading order low energy constant $F_0$ representing 
the pseudo-scalar decay constant in the SU(3) chiral limit (table~\ref{tab:F0})
have been surprisingly low.

RBC-UKQCD found \cite{Allton:2008pn}
that NLO SU(3) formula does not describe pseudoscalar masses
and decay constants above $m_\pi \simeq 400{\rm MeV}$. In addition to surprisingly low results 
for the LEC, RBC-UKQCD obtained a slightly low result $f_\pi = 124.1(3.6)(6.9)$ MeV ; with a single
lattice spacing this is within systematic error of the physical value.
At simulated masses and even at the pion mass the SU(3) correction to leading order was surprisingly large,
figure~\ref{fig:RBCUKQCDFPI}.
For this reason, they introduced the application of $SU(2)$ \tchpt to 
both pionic and kaonic quantities in an underlying
2+1 flavor simulation \cite{Antonio:2007pb,Allton:2008pn}, and this has been adopted by other groups.

In this SU(2) effective theory the kaon is treated as a heavy external
meson field \cite{Antonio:2007pb,Gasser:2007sg,Allton:2008pn}. The least convergent terms expand in 
$\left(\frac{m\pi}{m_K}\right)^2$, and this appears the more convergent 
approach for current data than $SU(3)$ expansion where the least convergent
terms expand in $\left(\frac{m_\eta}{4 \pi f}\right)^2$.

\begin{figure}[hbt]
\begin{minipage}{.5\textwidth}
\includegraphics[width=\textwidth]{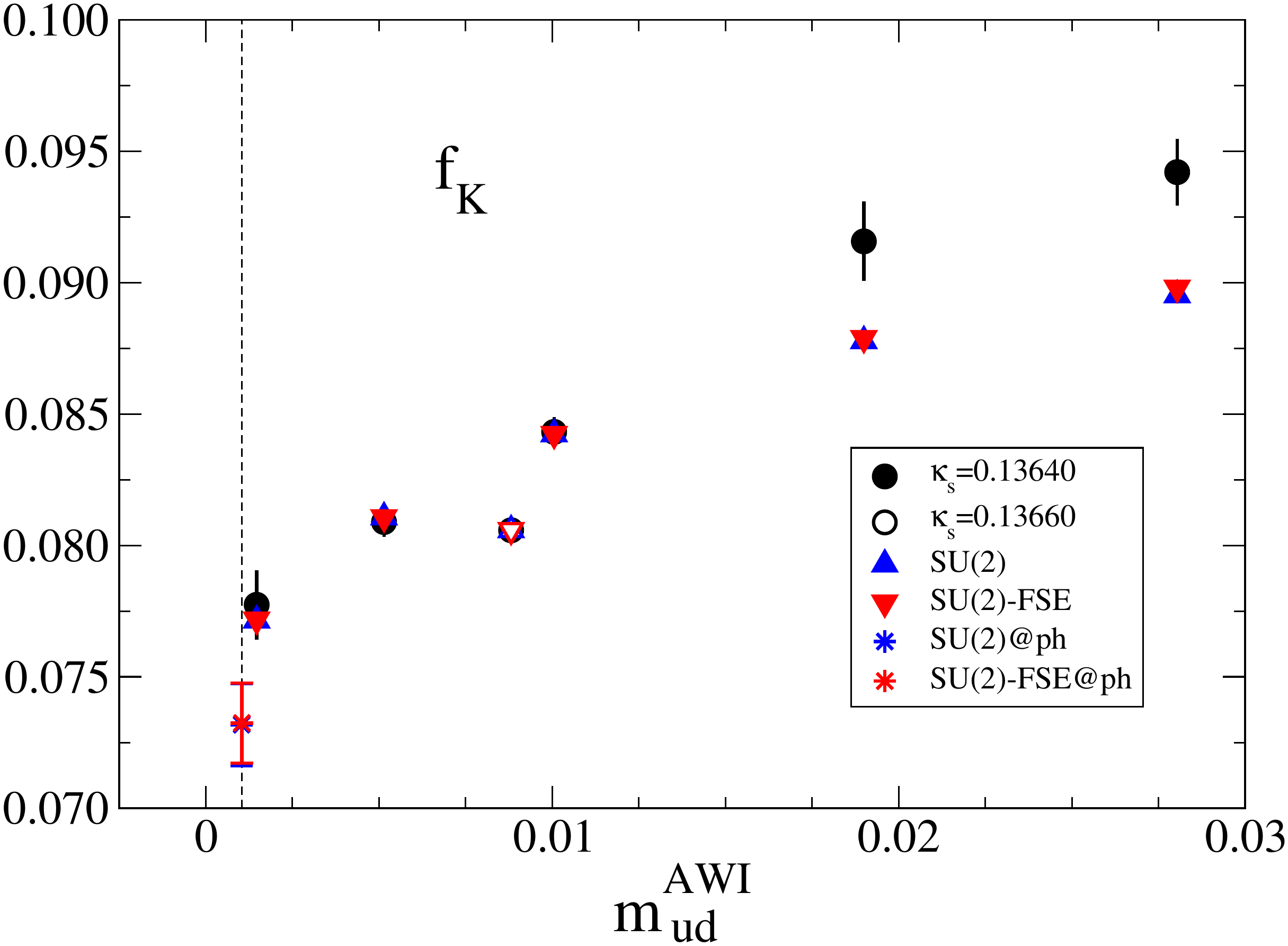}
\end{minipage}
\begin{minipage}{.5\textwidth}
\includegraphics[width=\textwidth]{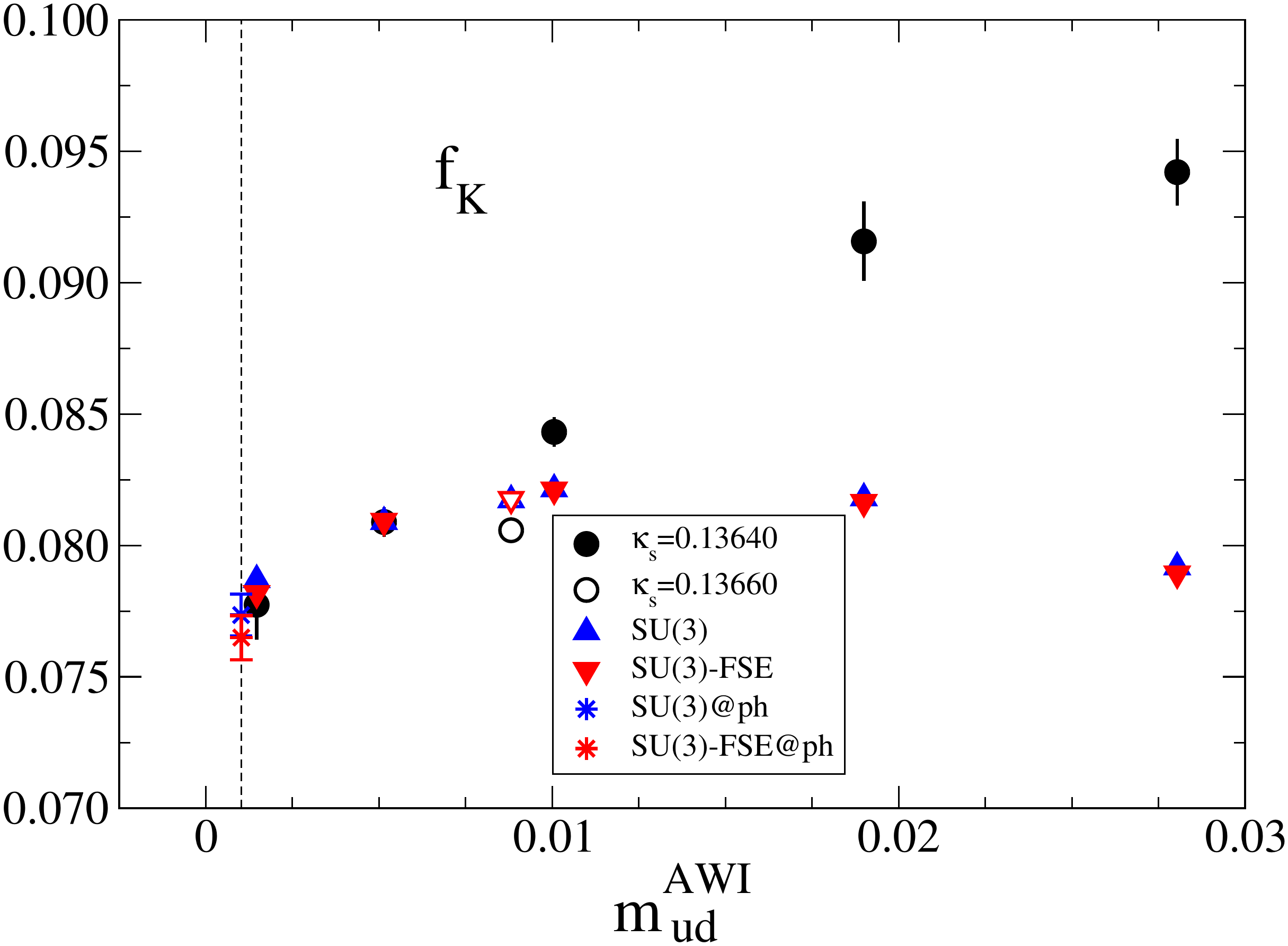}
\end{minipage}
\caption{PACS-CS NLO fits to the kaon decay constant using SU(2) (left) and SU(3) (right) \tchpt.
As with RBC-UKQCD SU(2) is found to give better agreement at larger masses.
The vertical dashed line represents the physical point, albeit in a small volume.
The ``FSE'' fits are corrected for finite size effects using CDH. SCALE?
\label{fig:PACSCSFK}
}
\end{figure}

The PACS-CS collaboration also found that kaon data was better
described by SU(2)\cite{Kuramashi:2008tb,Kadoh:2008sq}, figure~\ref{fig:PACSCSFK}; however their result for $F_0$ was
much larger and the conclusion about simulated 200-400 MeV pions differs. 
The underlying data for $f_\pi$ is qualitatively similar between PACS-CS and
RBC-UKQCD but differs by around 4\%. PACS-CS determine $Z_A$ from one loop
lattice perturbation theory; 
whether both  the overall scale and curvature of their chiral fits
would become closer with a non-perturbative $Z_A$ remains an open question.

The MILC collaboration found a larger $F_0$ than RBC-UKQCD \cite{Bazavov:2009bb}; 
however this is not necessarily 
discrepant when one considers that the RBC-UKQCD results were from a single lattice spacing.
The SU(3) chiral limit could be clarified relatively easily by simulations with 
artificially light ``strange'' quarks..

\begin{table}[hbt]
\begin{center}
\begin{tabular}{lll}
\hline
Collaboration & $F_0$ (MeV) & $F/F_0$ \\
\hline

{\footnotesize{MILC}} \cite{Bazavov:2009bb}   & 106(8)  & 1.15(5)$\binom{+13}{-3}$ \\
{\footnotesize{RBC/UKQCD}} \cite{Allton:2008pn}& 93.5(7.3)  & 1.229(59) \\
{\footnotesize{PACS-CS}}\cite{Aoki:2008sm}  & 118.5(9)   & 1.078(58)
\\
\hline
\end{tabular}
\end{center}
\caption{
\label{tab:F0}
There remains some spread in recent results for the leading order low energy constant $F_0$ of the
SU(3) chiral effective theory. A low value would imply SU(3) corrections for kaons in particular
must be uncomfortably large. $F$ is the SU(2) low energy constant and here the normalisation is such that
the pion decay constant is around 131 MeV.
}
\end{table}

\subsection{The curious linearity of lattice data}

BMW have not yet published a detailed comparison of their analytic fits to
their chiral fits \cite{Durr:2008zz}. However several collaborations have found suprising linearity in their data
for $f_\pi$.

\begin{figure}[hbt]
\begin{minipage}{0.5\textwidth}
\includegraphics[width=0.7\textwidth]{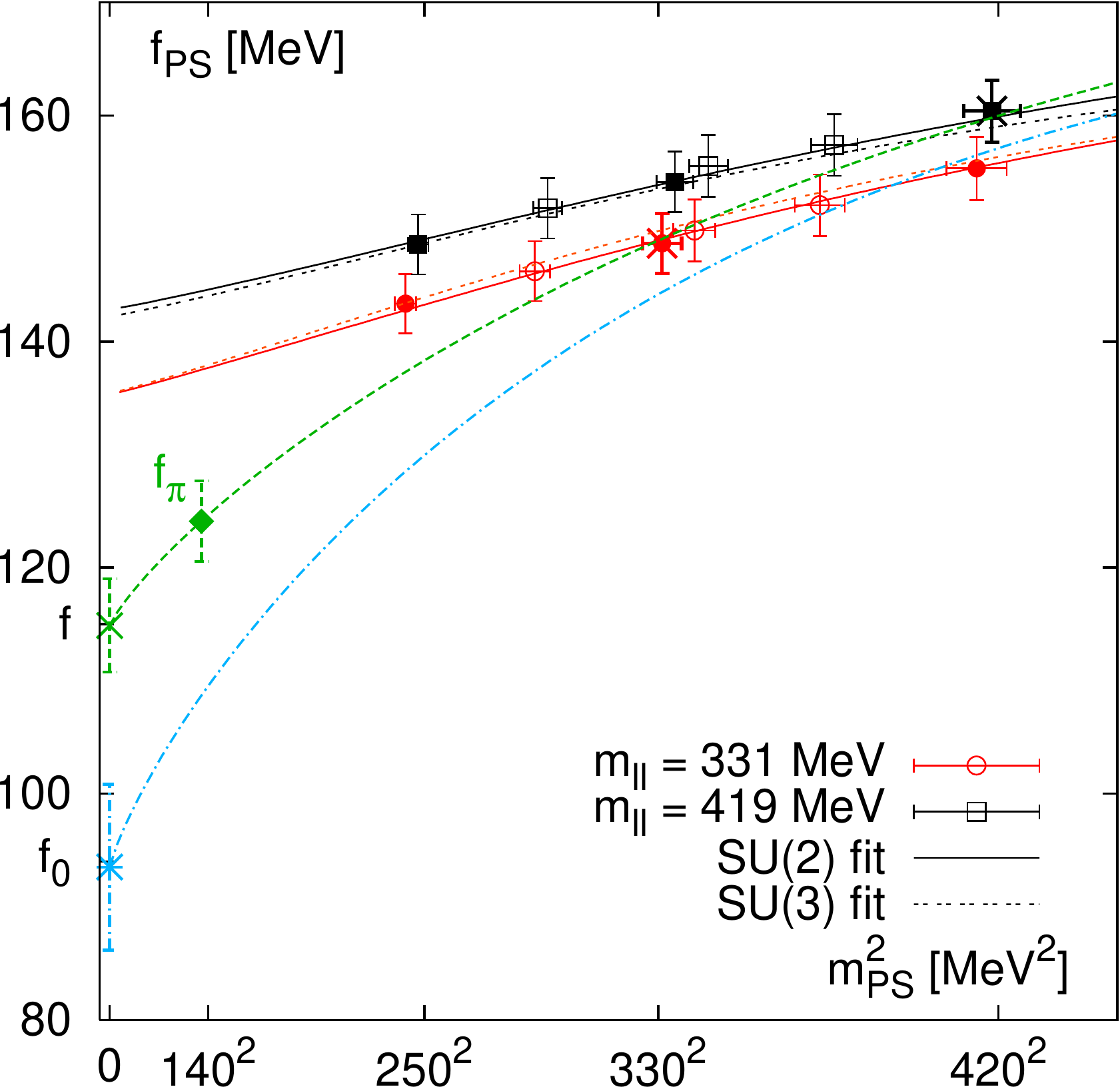}
\end{minipage}
\begin{minipage}{0.5\textwidth}
\includegraphics[width=\textwidth]{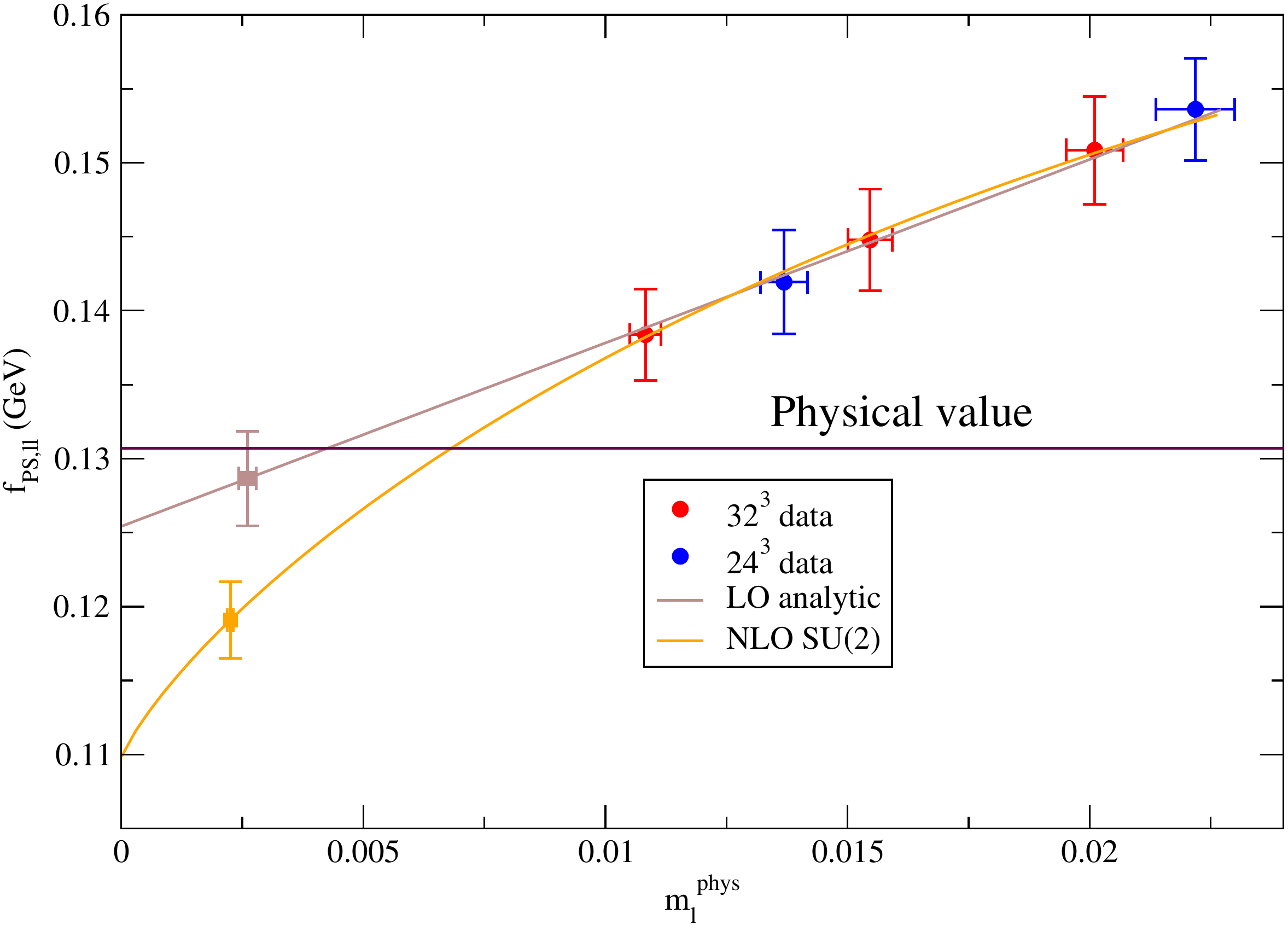}
\end{minipage}
\caption{
\label{fig:RBCUKQCDFPI}
\label{fig:ChrisFpi}
Left: NLO chiral extrapolation of RBC-UKQCD's dynamical DWF decay constant for $f_\pi$ on their 1.73GeV ensemble.
The decay constant in the SU(3) limit is surprisingly low, and corrections in this expansion are uncomfortably 
large. This plot displays partially quenched data. 
Right: Since the Kaon conference RBC-UKQCD have presented a preliminary continuum limit for their
simulations. The data points here are adjusted to represent values in the continuum limit and compared
to both an NLO and linear extrapolation. The scale is taken from $M_\Omega$, and NLO \tchpt yields
a pion decay constant that is inconsistent with its physical value, while linear extrapolation is in agreement.
The inconsistency is certainly explicable by a naturally sized NNLO effect, however the coincidence
of cancellation between orders to leave such linear behaviour at this level of precision is curious.
This plot displays only the unitary datapoints for simplicity.
}
\end{figure}

Since the Kaon conference RBC-UKQCD have presented a second lattice spacing \cite{Kelly,RDM}
at Lattice 2009, where they found good scaling compared with their earlier calculation and 
form conclusions about the continuum limit with theoretically pure dynamical chiral fermions.
They set the scale using the $\Omega$ baryon. 
Figure~\ref{fig:ChrisFpi} demonstrates that while a simple linear from their 290-450MeV
simulated masses produces the PDG $f_\pi$, NLO extrapolation does not.
The discrepancy  is consistent with a naturally sized NNLO effect; however the 
cancellation to produce linearity within statistical error is striking.
This data corroborates the BMW claim that Taylor expansion in the region directly above the 
physical masses shows good convergence; however the presence of known non-analytic terms makes it 
difficult to continue analytic fits to lighter masses and RBC-UKQCD used
the difference between analytic and chiral fits as a systematic error in predictions\cite{Kelly}.

\subsection{Low energy constants}

Lattice determinations of the leading order and next-to-leading order LEC's have been surveyed 
in detail by S. Necco, and by FLAG in excellent recent reviews \cite{Necco:2009cq,FLAG}. I reproduce
here some of the SU(2) NLO summary in table~\ref{tab:NeccoF}, as I believe this
is more convergent and hence more meaningful than SU(3). 
I recommend arXiv:0901.4257 \cite{Necco:2009cq} for details and further results, including a survey 
of the large body of $\epsilon$-regime results which are not covered
in this paper.

\begin{table}[hbt]
\footnotesize{
\begin{tabular}{llllll}
\hline
& & & & & \\[-0.2cm]
Collaboration & $N_f$  & $F$ (MeV) & $F_0$ (MeV) &
$\Sigma^{1/3}$ (MeV)  & $\Sigma_0^{1/3}$ (MeV)\\
& & & & & \\[-0.2cm]
\hline
& & & & & \\[-0.2cm]
{\footnotesize{ETM}
\cite{Dimopoulos:2008sy}
}   & 2   &  86.03(5) & & 267(2)(9)(4)
& \\
{\footnotesize{JLQCD/TWQCD}\cite{Noaki:2008iy}} & 2 &
79.0(2.5)(0.7)$\binom{+4.2}{-0.0}$  & &
235.7(5.0)(2.0)$\binom{+12.7}{-0.0}$ & \\
{\footnotesize{MILC}\cite{Bernard:2007ps}} & 2+1 &  & &$278(1)\binom{+2}{-3}(5)$  & $242(9)\binom{+5}{-17}(4)$\\
{\footnotesize{RBC/UKQCD}\cite{Allton:2008pn}}  & 2+1 & 81.2(2.9)(5.7) & & 255(8)(8)(13)  & \\
{\footnotesize{PACS-CS}\cite{Aoki:2008sm}}  & 2+1  & 90.3(3.6) & 83.8(6.4) &309(7)  & 290(15) \\
{\footnotesize{ETM}\cite{Frezzotti:2008dr}}  & 2   &  86.6(4)(7) & & 264(2)(5)& \\
\hline
\end{tabular}
}
\footnotesize{
\begin{tabular}{lllllll}
\hline
& & & & & &\\[-0.4cm]
Collaboration & $N_f$ & $\bar{l}_3$ (SU(2)) & $\bar{l}_4$ (SU(2)) & $\bar{l}_3$ (SU(3)) &   $\bar{l}_4$ (SU(3))  \\
\hline
{\footnotesize{ETM}}   & 2 & 3.42(8)(10)(27)
& 4.59(4)(2)(13) & &  \\ 
{\footnotesize{ETM}}  & 2 & 3.2(4)(2)
& 4.4(1)(1) & & \\
 {\footnotesize{JLQCD/TWQCD}}  & 2 & 3.44(57)$\binom{+0}{-68}\binom{+32}{-0}$ &
4.14(26)$\binom{+49}{-0}\binom{+32}{-0}$ & & \\[0.1cm]
 {\footnotesize{MILC}}  & 2+1  &  & &
$1.1(6)\binom{+1.0}{-1.5}$ & $4.4(4)\binom{+4}{-1}$ &\\
 {\footnotesize{RBC/UKQCD}}  & 2+1 & 3.13(33)(24) & 4.43(14)(77) &
2.87(28) & 4.10(5)  \\
 {\footnotesize{PACS-CS}}  & 2+1 &  3.14(23) & 4.04(19) & 3.47(11) & 4.21(11) \\ 
\hline
\end{tabular}
\caption{
\label{tab:NeccoF}
\label{tab:NeccoLs}
(Abridged) summary tables from \cite{Necco:2009cq}, which also reviews SU(3) LEC's and the
large body of work making use of the $\epsilon$-regime.
Here F is in the convention where $f_\pi\simeq 92 $MeV
}
}
\end{table}

\subsection{Pion form factors and $l_6$}
The pion vector form factor is related to the low energy constant $l_6$:

$$ \langle \pi^+(p^\prime)|V_\mu|\pi^+(p)\rangle = F^{\pi\pi}(q^2) (p_\mu+p^\prime_\mu) $$
$$
\langle r_\pi^2 \rangle = 6 \frac{d}{dq^2} f^{\pi\pi}(q^2)|_{q^2=0}
 =
-\frac{12 l_6^r}{f^2} - \frac{1}{8\pi^2 f^2}\left(\log\frac{m_\pi^2}{\mu^2} + 1
 \right) .
$$
A summary of recent results for the pion charge
radius is given in table~\ref{tab:piff}. There are presently
considerable discrepancies in the results at intermediate masses prior
to chiral extrapolation, and consequently differing conclusions.

\begin{figure}[hbt]
\includegraphics[width=0.5\textwidth]{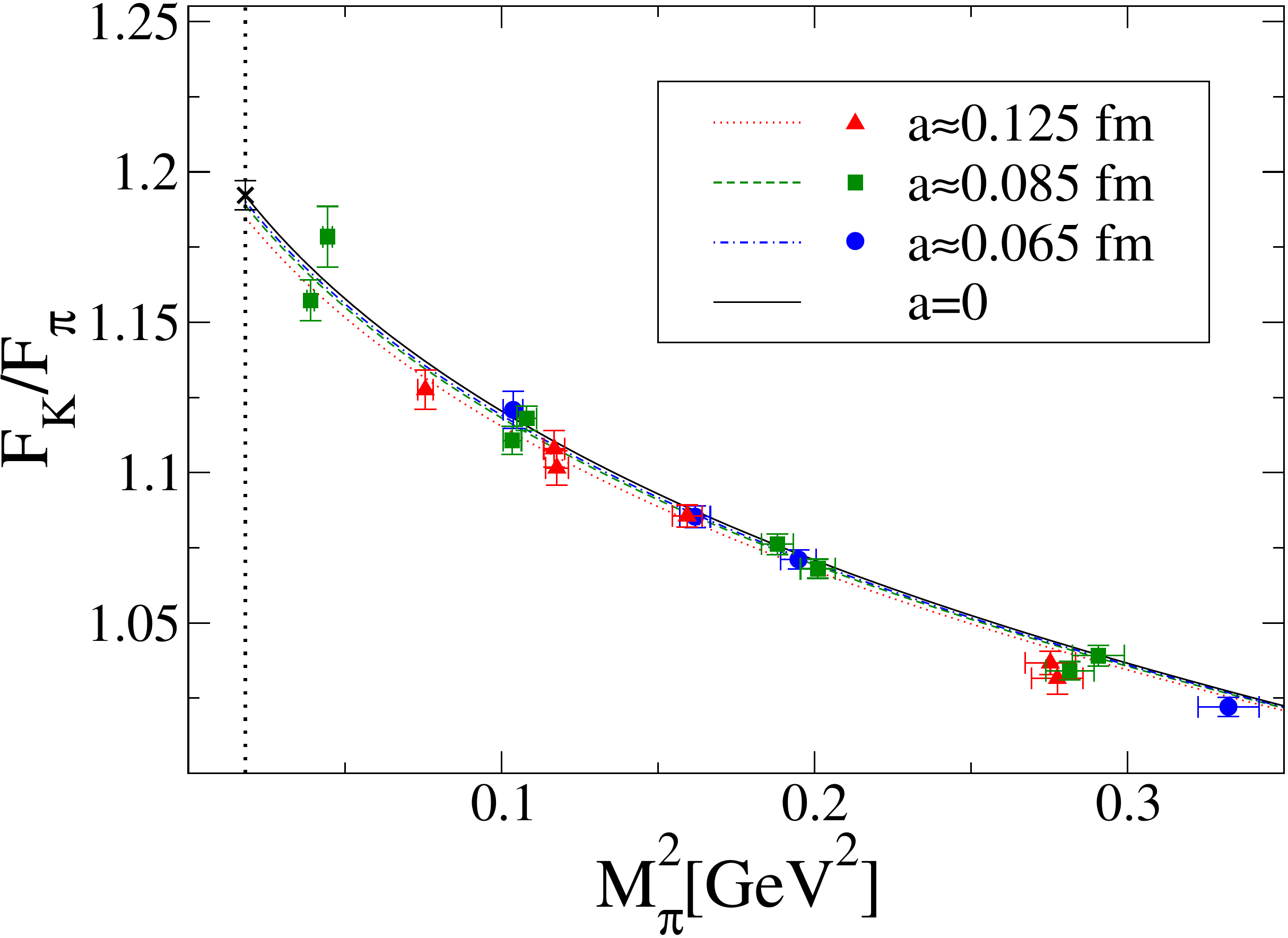}
\includegraphics[width=.45\textwidth]{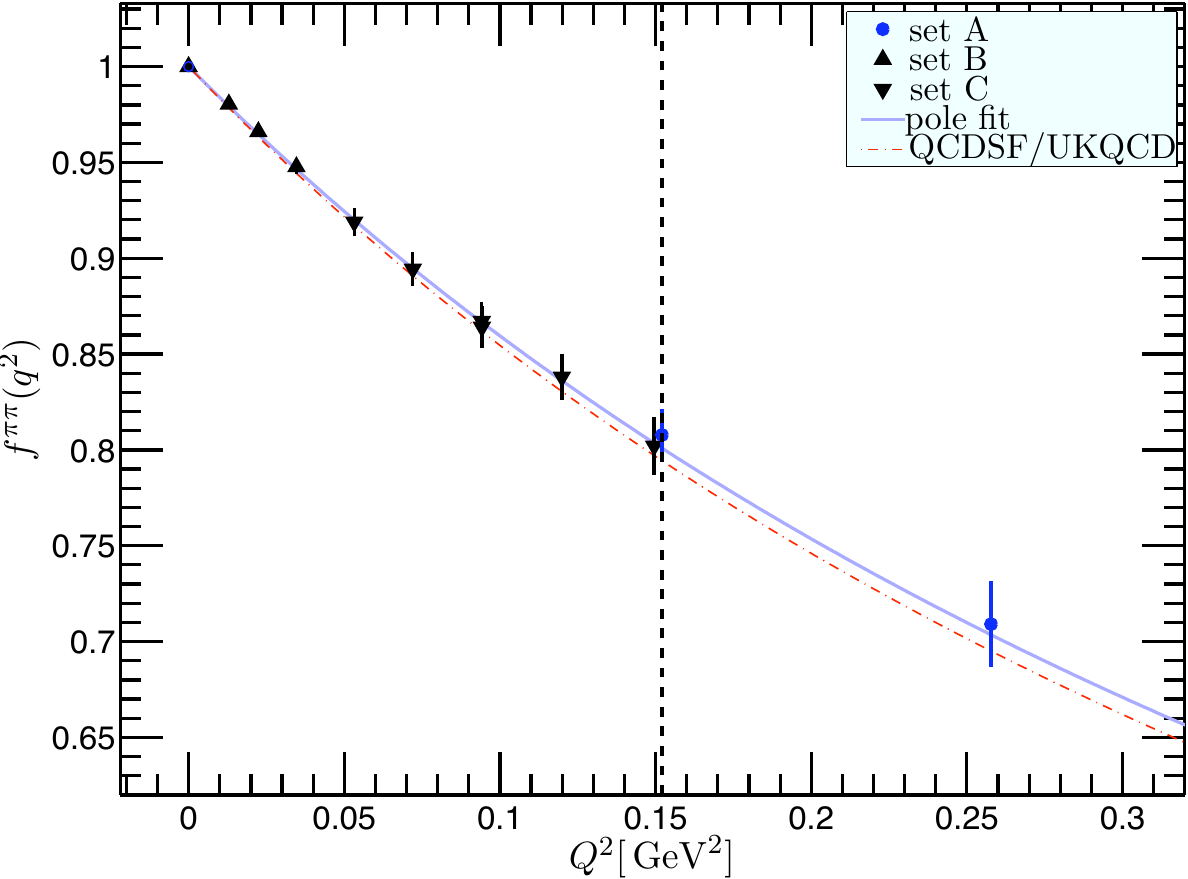}
\caption{
\label{fig:FKFPI}
\label{fig:piFF}
LEFT: BMW's chiral extrapolation of $f_K/f_\pi$ \cite{Durr:2008zz}. This includes an impressive
190MeV lightest simulated mass, 4 fm volumes and a continuum limit. Detailed exposition in 
print is absent, but nevertheless this
is an exciting calculation for which further details are eagerly awaited.
RIGHT: 
RBC-UKQCD mapped out the deep infra-red region of the pion form factor using twisted boundary conditions. ETMC used a similar approach down $0.05 \mathrm{GeV}^2$
\cite{Boyle:2008yd}
}
\end{figure}

ETMC and JLQCD \cite{Frezzotti:2008dr,Kaneko:2008kx,Aoki:2009qn,Hashimoto:2005am}
obtain somewhat lower values than RBC-UKQCD and QCDSF
in the region $300 \le m_\pi \le 500$ MeV, and conclude the NNLO effects
in the chiral expansion are significant. This calculation
typically involves the use of a number discrete Fourier modes
for $q$ with model dependent interpolation used to obtain the 
derivative at $q^2=0$.
RBC-UKQCD and ETMC \cite{Boyle:2008yd,Frezzotti:2008dr}
obtained good resolution in the low $q^2$ region, figure~\ref{fig:piFF}.
Here, a twisted boundary condition with phase $e^{i\theta}$ applied to valence
Fermions on the simulated torus allows to vary lattice 
momenta smoothly between the 
standard periodic and anti-periodic Fourier modes, enabling several
momenta between the zero and first Fourier modes
\cite{Boyle:2007wg}. 
RBC-UKQCD found that NLO
\tchpt yielded good results for the pion charge radius, but this 
remains a puzzle however, since ETMC also used
twisted boundary conditions but concluded NNLO \tchpt was required.

The determination of the pion scalar charge radius by JLQCD is worth noting
and which has been performed including the effects of disconnected
quark flow diagrams for the first time\cite{Aoki:2009qn}.

\begin{table}[hbt]
\begin{center}
{\footnotesize{
\begin{tabular}{c|c|c|c}
Collab & Action& $\langle r_\pi^2 \rangle_V$ & $\langle r_\pi^2 \rangle_S$\\
\hline
ETMC  \cite{Frezzotti:2008dr}     & 2f TM         & 0.456(38)  & \\
JLQCD \cite{Kaneko:2008kx,Aoki:2009qn}    & 2f Overlap    & 0.409(23)(37)  & 0.617(79)(66)\\
JLQCD \cite{Hashimoto:2005am}  & 2f Clover     & 0.396(10)  & 0.60(15)\\
QCDSF \cite{Brommel:2006ww}    & 2f Clover     & 0.441(19) &  \\
\hline
RBC-UKQCD\cite{Boyle:2008yd}  & 2+1f DWF      & 0.418(31) & \\
LHPC     \cite{Bonnet:2004fr} & 2+1f MILC/DWF & 0.310(46) & 
\end{tabular}
}
}
\end{center}
\caption{Summary of recent results for the pion charge radius from lattice QCD
\label{tab:piff}
}
\end{table}

\subsection{$L_{10}^r$}

Here we see the first of many quantities that are made accessible with chirally symmetric lattice
actions. $L_{10}^r$ is related to the electromagnetic component of the splitting between $\pi^+$ and $\pi^0$,
and is determined from small difference between the axial and vector vacuum polarisation functions.

\begin{eqnarray}
\Pi_{V_{\mu\nu}} -\Pi_{A_{\mu\nu}} &=& \left(q^2 \delta_{\mu\nu} - q_\mu q_\nu \right)
\nonumber                  \Pi_{V-A}^{(1)} - q_\mu q_\nu \Pi^{(0)}_{V-A} \\
\nonumber
\Pi^{(1)}_{V-A}&=&
 -\frac{f_\pi^2}{q^2} - 8 L_{\rm 10}^r(\mu) 
 -\frac{\log(\frac{m_\pi^2}{\mu^2}+\frac{1}{3} - H(x))}{24\pi^2}
\end{eqnarray}
JLQCD produced a beautiful first determination in their $n_f=2$ overlap simulation
\cite{Shintani:2008qe}.
obtaining the pion e-m mass squared splitting (in chiral limit; c.f. PDG with physical $u,d$ masses $1261 \mathrm{MeV}^2$):

$$
L_{10}^r(m_\rho) = -5.2(2)^{+5}_{-3} \times 10^{-3}
$$
$$
m_{\pi^\pm}^2 - m^2_{\pi^0} = 993(12)(^{+0}_{-135})(149) \mathrm{MeV}^2
$$

A second calculation has been performed by RBC-UKQCD with 2+1f domain wall fermions \cite{Jan}.
The calculation is also interesting in that it produces information about the radius of convergence of
\tchpt in momentum space, in addition to its convergence in pion mass; these are similar.

\section{$V_{us}$ }

\label{Sec:Vus}
The two best lattice constraints for $V_{us}$ involve SU(3) breaking effects:
for $\frac{f_K}{f_\pi}$ this is O(20\%), and for $f_0^{K\pi}(q^2=0)$ 
this is O(4\%).  While the $K_{l3}$ approach is less mature, it
looks very promising and is rapidly becoming better studied.
We consider these two quantities in turn.

\subsection{$V_{us}$ from $f_K/f_\pi$}
\label{Sec:FKFPI}

This topic was reviewed in detail by Lellouch \cite{Lellouch:2009fg}.
The calculation with the smallest quoted errors is the mixed action
HISQ/Asqtad staggered calculation by HPQCD\cite{Follana:2007uv}, followed by
the smeared clover simulation from BMW\cite{Durr:2008zz}.

The HPQCD calculation used a mixed action. While published in an abbreviated 
form, it appears to have not followed MILC's best practice of performing a
partially quenched rs-\tchpt analysis to absorb unitarity violations. Rather, 
it appears to have matched the HISQ valence pseudoscalar 
Goldstone taste in to the (irrelevant here) 
AsqTad valence pseudoscalar Goldstone taste. This differs from the 
various relevant masses of sea pseudoscalar multiplet. 
It is hard for lattice theorists
not directly involved in the calculation for form a clear judgement with the available
information.

Figure~\ref{fig:FKFPI} (left) reproduces the truly impressive BMW simulation
that spans three lattice spacings and includes volumes up to 4 fm and 
pion masses down to an impressive 190MeV. The BMW calculation has not yet
received a detailed exposition in print, and has so far lacked a non-perturbative
determination of the axial current renormalisation so important checks on 
$f_K$ and $f_\pi$ seperately have not yet been performed; it seems likely
non-perturbative operator improvement was not used and resulted in their
dual use of $a$ and $a^2$ continuum extrapolation.
Their chiral extrapolation was interesting but details are sparse; 
chiral expansion and Taylor expansion fits 
were combined weighted by quality of fit and only the combined fit has been
shown. Publication of the qualities of fit, 
and direct comparison of the two in the style of figure~\ref{fig:ChrisFpi},
would be interesting since continuing an analytic fit downwards is not without risk and it is 
not yet clear how much the lightest data points constrain chiral curvature.

Since the Lellouch review, ETMC have 
updated \cite{Blossier:2009bx} their $N_f=2$ twisted mass result to $f_K/f_\pi = 1.210(18)$.
As this will not substantially change the lattice average I recommend continued
use of that quoted by Lellouch \cite{Lellouch:2009fg} while noting the leading publications
are in formats that are not amenable to critical assessment
$$
f_K/f_\pi = 1.194(3)(10).
$$

\subsection{$V_{us}$ from $K_{l3}$}
\label{Sec:KL3}

The semi-leptonic form factor 
$$ \langle \pi(p^\prime)|V_\mu|K(p)\rangle 
= 
f_+(q^2) (p_\mu+p^\prime_\mu) + f_-(q^2) (p_\mu-p^\prime_\mu) 
$$
can be computed precisely using several ``double ratios'' following 
\cite{Becirevic:2004ya,Hashimoto:1999yp}, such as
$$
\frac{\langle K(\vec{0}) | V_0 | \pi(\vec{0}) \rangle \langle K(\vec{0}) | V_0 | \pi(\vec{0}) \rangle}
     {\langle K(\vec{0}) | V_0 | K(\vec{0}) \rangle \langle \pi(\vec{0}) | V_0 | \pi(\vec{0}) \rangle}
= \frac{(m_K+m_\pi)^2}{4m_K m_\pi} |f_0(q^2_{\rm max})|^2
$$

\begin{table}[hbt]
\begin{tabular}{lllcccc}
\hline
\hline
Ref. & $N_f$ & action & $a[\fm]$ & $L[\fm]$ & 
\begin{tabular}{c}$M_\pi^{min}[\mev]$\\[-0mm]
typ/val\end{tabular} & $f_+(0)$\\
\hline
{JLQCD} \cite{Tsutsui:2005cj}& 2 & NP SW & 0.09 & 1.8 & 550/550 & 0.967(6)\\
{RBC}\cite{Dawson:2006qc}    & 2 & DWF & 0.12 & 2.5 & 490/490 & 0.968(9)(6)\\
{ETMC} \cite{Lubicz:2009ht}  & 2 & tmQCD & 0.11 & 2.7 & 260/260 & 0.9581(57)(35)\\
{FNAL/MILC}\cite{Okamoto:2004df} & 2+1 & KS+Wil &  &  & & 0.962(6)(9)\\
{RBC/UKQCD}\cite{Boyle:2007qe} & 2+1 & DWF & 0.11 & 1.8, 2.8 & 290/240 & 0.9644(33)(34)(14)\\
\hline
\hline
\end{tabular}
\caption{An update of the corresponding table by Lellouch \cite{Lellouch:2009fg} to include new ETMC results
\label{tab:KL3}
}
\end{table}

\begin{figure}[hbt]
\includegraphics*[width=.5\textwidth]{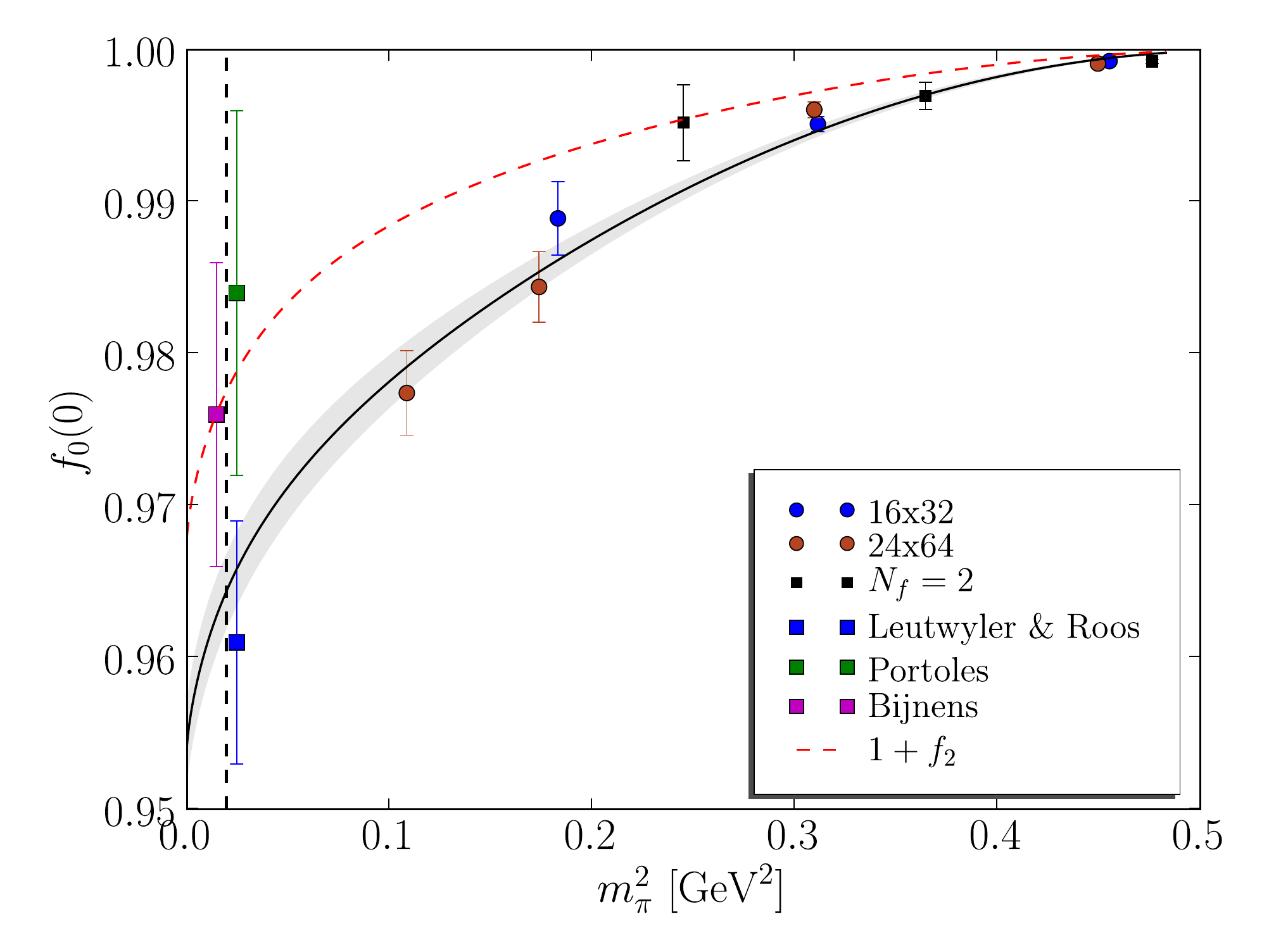}
\includegraphics*[width=.45\textwidth]{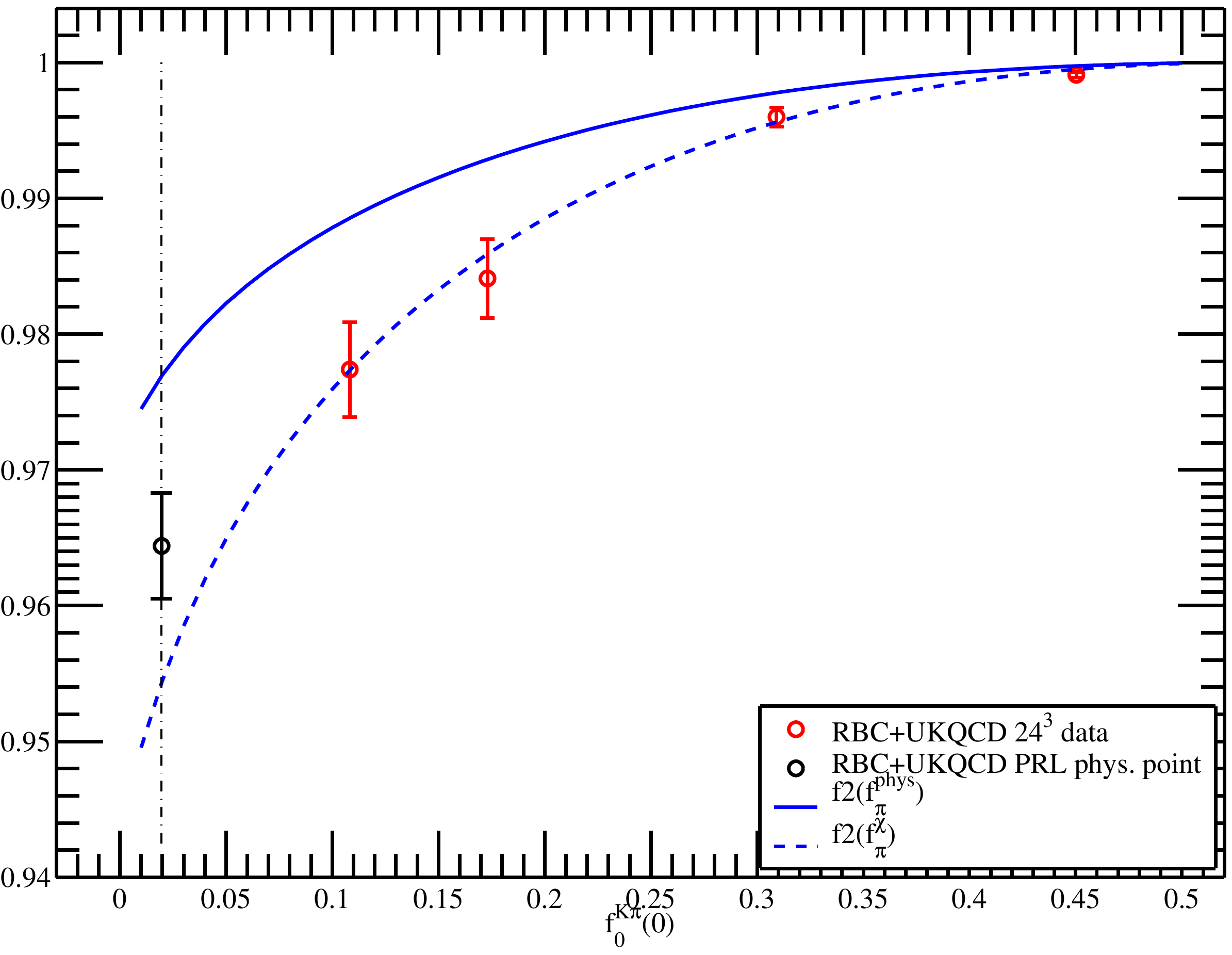}
\caption{LEFT: RBC-UKQCD chiral extrapolation of $f_+(0)$ represents the best lattice constraint and has
2+1 flavors. RIGHT: Courtesy A. J\"uttner -- using RBC-UKQCD's result for $F_0$ to better define $f_2$ 
results in NLO describing the data well and the chiral expansion of this small breaking looking naturally
convergent. A redefined $\Delta_f$ would be near zero.
\label{fig:KL3}
\label{fig:KL3f0}
}
\end{figure}
The form factor $f_+(q^2=0)$ is of key interest for $V_{us}$ and has chiral expansion $1 + f_2 + f_4 \ldots $ 
where $f_n \simeq O( M_{K,\pi}^n / (4 \pi F_0)^n )$.
The best constraint in, table~\ref{tab:KL3} uses 2+1 flavors and is from RBC-UKQCD \cite{Boyle:2007qe}. Their
chiral extrapolation is shown in figure~\ref{fig:KL3} obtaining 
$$
f_+(0)=0.9644(33)^{\rm stat}(34)^{\rm extrapolation}(14)^{\rm disc}.
$$
The calculation used Fourier modes, with model dependence in $q^2$ interpolation 
and chiral extrapolation; there was no controlled continuum limit and a discretisation systematic
error was budgeted at 4\% of $1-f_+(0)$.
The historical use of $f_\pi$ as the denominator for $f_2$ is not the only
sensible choice. 
Taking the above RBC-UKQCD result for $F_0$ in table~\ref{tab:F0}
better matches the data and rearranges the expansion at NNLO and above.
Different reasonable choices of $f_2$ seem to be an important systematic
in SU(3) based extrapolations of the form factor.
RBC-UKQCD presented additional data at the lattice conference making use
of twisted boundary conditions to simulate directly at $q^2=0$ 
\cite{Flynn:2008hd,zanotti}
and have developed an SU(2) formalism  \cite{Flynn:2008tg} which they will
use in the chiral extrapolation of their new data.

The ETMC $n_f=2$ calculation \cite{Lubicz:2009ht} had a notably robust chiral extrapolation
with pion masses as low as 260 MeV and made use of both SU(3) 
and SU(2) formalisms in their chiral extrapolation. An adjustment
was made (but not quoted here) to ``correct'' the lattice calculation for 
the missing strange quark using the SU(3)
expression for $f_2$. The degree to which leading effects of the absent strange loops 
are already reabsorbed when the lattice spacing is determined makes this unconvincing; 
I prefer to quote the unadjusted ETMC calculation as an excellent $n_f=2$ result.


\section{$B_K$}
\label{Sec:BK}
The matrix element
$$
 B_K = \frac{\langle K^0 | {\cal{O}}_{\rm VV+AA}|\bar{K}^0 \rangle}
              {\frac{8}{3} \langle K^0| A_0\rangle\langle A_0|K^0\rangle}
$$
is multiplicatively renormalised in the continuum and for lattice actions with chiral symmetry.
Non-symmetric actions must deal with unphysical taste/chirality mixings which inflate errors.
In some cases \emph{mixed actions} are used (e.g. ALV, ETMC), where a valence quark action with
multiplicative operator renormalisation is combined with a cheaper action for sea quarks; 
resultant unitarity violations should be fitted away, typically using mixed action chiral perturbation theory.

\subsection{$B_K$ in the quenched approximation}
For many years the benchmark quenched $B_K$ calculation was from JLQCD using many
lattice spacings with the staggered fermion action 
\cite{Aoki:1997nr}.
This calculation involved the lattice perturbative treatment of a taste mixed operator basis.
The continuum limit of $$B_K^{\overline{\mathrm{MS}}}(n_f=0,2\mathrm{GeV})=0.565(4)(5),$$
obtained by JLQCD \cite{Nakamura:2008xz} using the chirally symmetric
domain wall fermion formulation, gives continuum limit consistent with various earlier
calculations. This includes RBC who used a different gauge action thus displaying universality.
JLQCD's results are particularly clean in that they apply non-perturbative
step scaling to enable matching to continuum perturbation theory at large scales.

This is substantially below the previous quenched ``benchmark'', and resolved a puzzle
highlighted by Lellouch. It also highlights that lattice renormalisation issues
of various formulations can be a very important effect.

\begin{figure}[hbt]
\begin{center}
\includegraphics*[width=0.48\textwidth]{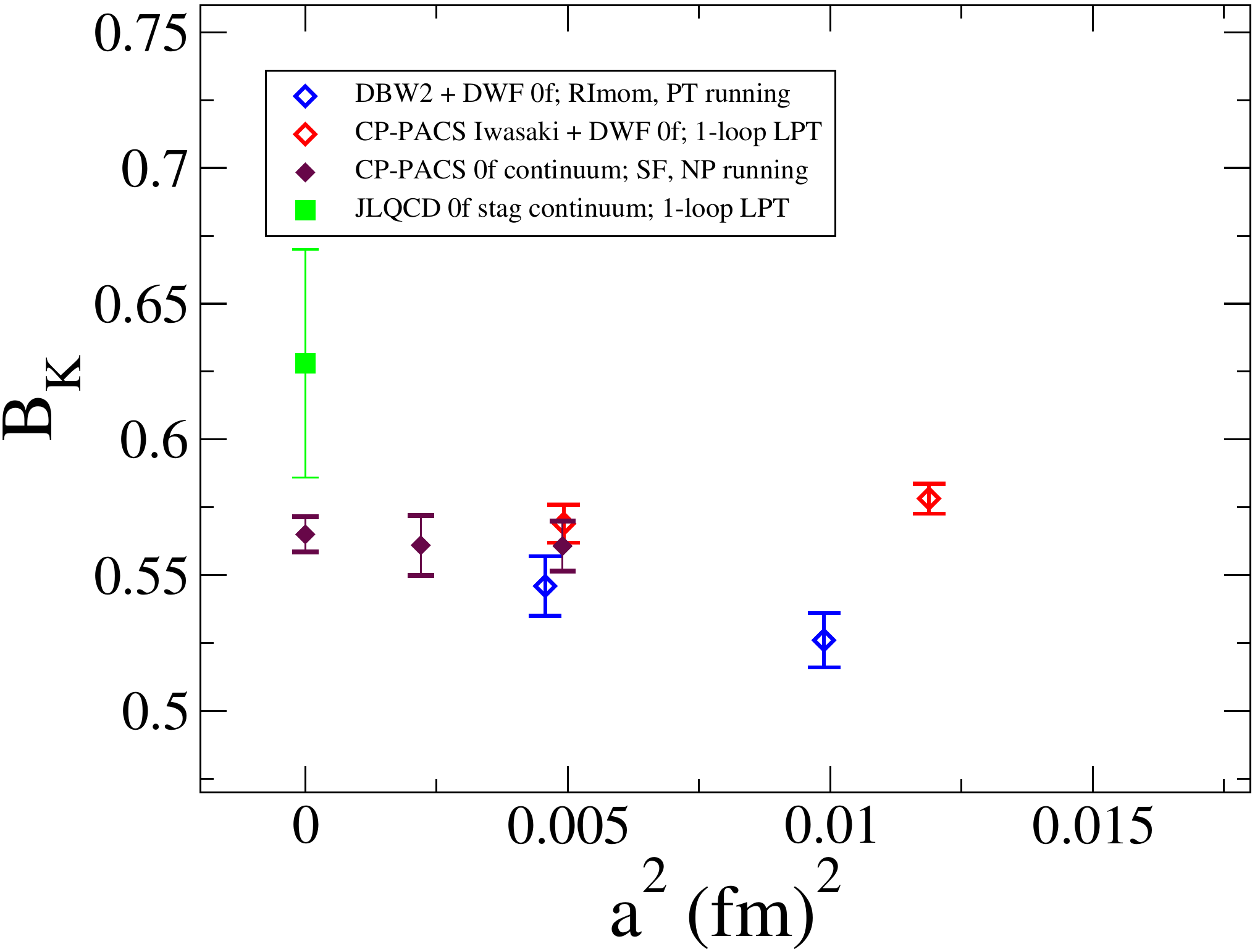}
\includegraphics*[width=0.48\textwidth]{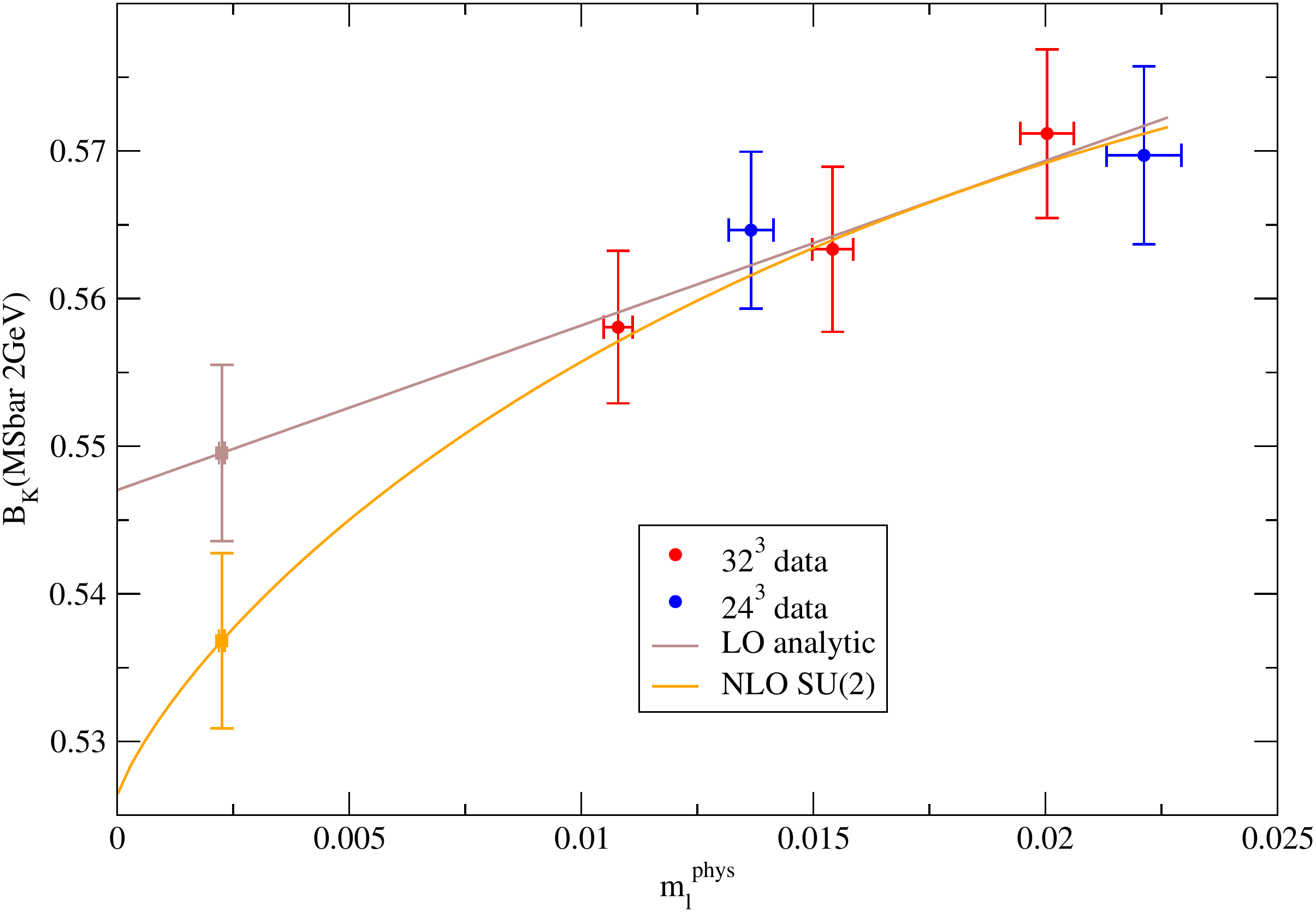}
\end{center}
\caption{LEFT: Quenched continuum limit of $B_K$ has been substantially clarified, 
with the previous benchmark using staggered fermions appearing suspect. The
chirally symmetric DWF formulation has only multiplicative renormalisation and CP-PACS
have taken a high quality continuum limit with non-perturbative renormalisation. 
This is plausibly consistent with both RBC (non-perturbative) and earlier (perturbatively renormalised) CP-PACS
results. RIGHT: RBC-UKQCD chiral extrapolation of $B_K$ in the continuum limit; data points are shifted
to the continuum limit and overlayed. Systematic error taken from difference between linear and SU(2) NLO fits.
\label{fig:Bkquenched}
}
\end{figure}

\subsection{2f $B_K$}

There are two recent and high quality determinations of $B_K$ in the two flavour theory.
JLQCD \cite{Aoki:2008ss}
made an impressive calculation with 2f overlap action, while
ETMC used a somewhat less aesthetic mixed action approach with Osterwalder-Seiler valence fermions with
2f twisted mass sea fermions 
\cite{Dimopoulos:2008hb}.
JLQCD obtained $B_K^{\overline{\rm MS}} (n_f=2,2 {\rm GeV})= 0.537(4)(40)$ 
and ETMC obtained $B_K^{\overline{\rm MS}} (n_f=2,2 {\rm GeV})= 0.56(2)$ 
both with single lattice spacings. The ETMC calculaton is preliminary and has not
yet addressed important unitarity violation systematics and I quote:
$$B_K^{\overline{\rm MS}} (n_f=2,2{\rm GeV})= 0.537(4)(40).$$

\subsection{2+1f $B_K$}

Two 2+1f calculations were considered in detail at the Kaon conference:
RBC-UKQCD's dynamical domain wall simulation, and the Aubin/Laiho/Van de Water
mixed DWF valence and staggered (MILC) sea calculation. 
RBC-UKQCD's programme had previously published results from a single
lattice spacing \cite{Allton:2008pn,Antonio:2007pb}, and this result had been combined
by Lellouch with the 2+1f staggered result of Gamiz {\it et al.}
\cite{Gamiz:2006sq} in his world average. I choose \emph{not} to include this staggered result 
in an average in light of the quenched discrepancy above.
The RBC calculation had a preliminary update (since the Kaon conference) at
Lattice 2009 with a second lattice spacing and a preliminary joint chiral and continuum extrapolation.
These results are sufficiently significant that they merit inclusion in these proceedings.
This year Aubin, Laiho and Van de Water 
\cite{Aubin:2009jh}
used a mixed action approach
with valence domain wall fermions on 2+1f of staggered sea quarks.

A detailed comparison
of both calculations is interesting in that estimates of systematic errors are in fact 
the dominant source of uncertainty. The statistical errors in both calculations are subleading
and it is worth emphasizing the subjective nature of the systematic error estimates that
dominate the quoted errors. The most important differences arise from the estimate of the
renormalisation and chiral-continuum extrapolation errors.

ALV base their error on the non-perturbative renormalisation
on the difference from mean field improved 1-loop lattice perturbation theory.
However, RBC-UKQCD now use multiple non-exceptional momentum 
renormalisation schemes to gain control over their error estimate. 

Given figure~\ref{fig:ChrisFpi}, RBC-UKQCD estimated its chiral-continuum
extrapolation error based on the difference between linear and SU(2) NLO extrapolation, figure~\ref{fig:Bkquenched}.
ALV apply SU(3), which is in principle less 
convergent than SU(2), and estimate the corresponding 
error based on varying a subset of analytic NNLO 
terms included in the fit. They do not include taste violation effects beyond NLO or non-analytic terms beyond NLO. 
It is likely that the difference here lies in the approaches to error 
estimation.

The RBC-UKQCD result, however, is a preliminary conference submission and has not yet been given 
an estimate of finite volume effects, nor been published as even a preprint.
Until these systematic errors have been published in peer reviewed journals I recommend continued use of \cite{Antonio:2007pb}

$$
B_K^{\overline{\rm MS}} (n_f=2+1,2{\rm GeV})
= 0.524(30)
$$


\begin{table}[hbt]
\begin{center}
\begin{tabular}{cccc}
  & ALV & RBC-UKQCD(Lat09) & RBC-UKQCD(2007)\\
\hline
$B_K^{\overline{\rm MS}}(n_f=2+1,2{\rm GeV})$ & 0.527(6)(20) & 0.537(19) & 0.524(10)(28)\\
\hline
NP renormalisation        & 3.3\% & 2.4\% & 2\%\\
Chiral/cont extrapolation & 1.9\% & 2.4\% & 2\%, 4\%\\
statistical               & 1.2\% & 1.1\% & 1.9\%\\
Scale \& masses           & 0.8\% & 0  \% & 1\%\\
Finite volume             & 0.6\% & -     & 1\%\\
\hline
total (quad)& 4.0\% & 3.5\% & 5.7\%
\end{tabular}
\end{center}
\caption{Leading calculations of $B_K$ with 2+1 f. 
Systematic error estimates are dominant, and a detailed breakdown of the sources
error useful. RBC-UKQCD use a more 
convergent SU(2) approach to \tchpt while ALV
absorb taste mixings of heavy pseudoscalars with SU(3) \tchpt.
}
\end{table}

\subsection{Renormalisation of $B_K$}

The dominant systematic error in $B_K$ has become 
the QCD renormalisation.
The four recent high quality 2f and 2+1f simulations 
\cite{Aoki:2008ss,Dimopoulos:2008hb,Antonio:2007pb,Aubin:2009jh,Kelly} all make use of the
Rome-Southamptom RI-mom scheme to match between lattice and the $\overline{MS}$ regularisation
schemes via an intermediate physically defined MOM scheme.

By historical accident, the particular kinematic point selected was an exceptional
momentum point, and suffered avoidably from 
\cite{Aoki:2007xm} strong non-perturbative 
contamination. This paper demonstrated that a better choice is to 
use non-exceptional momenta as the intermediate MOM scheme, and preliminary results
have been presented at the lattice conference using a 1-loop calculation by Sachrajda
and Sturm. In fact, multiple ostensibly equivalent 1-loop schemes were used and
the spread gave a much more robust estimate of the systematic error at around 2.4\%,
leaving it as the (jointly) dominant source of error.

While other components of the lattice determination of $B_K$ will continue the
rapid progress seen recently, the determination of this matching to two loops at \emph{this}
non-exceptional momentum point is required for reduction of the lattice error on $B_K$
to below around 2\%.

The non-exceptional renormalisation point is also even more important to the
robust determination of the matrix elements of supersymmetry induced 
four quark operators\cite{Jan08}.

\section{Conclusions}

These are truly exciting times for lattice QCD as simulations push
steadily closer to the physical point in a large volume and in the
continuum limit. There has been tremendous recent progress in a number of important
areas; understanding of the chiral expansion of QCD has been 
greatly helped by Lattice QCD; Lattice QCD input to $V_{us}$ is hugely important
and the kaon bag parameter has been substantially better determined. 

The kaon is the ideal system for Lattice QCD; it contains rich
and non-trivial CP violating physics at energy scales that can be simulated 
without large cut-off effects. Nascent work on electromagnetic and
iso-spin breaking effects was beyond the scope of this review; however, it appears
that there are no barriers to the continued improvement in the precision of lattice
simulations beyond those posed by computing power.

\section{Acknowldgements}

PAB is funded by RCUK fellowship, and wishes to warmly thank both  KEK and
the CCS at the University of Tsukuba for both support and hospitality.


\end{document}